\documentclass{article}

\usepackage{microtype}
\usepackage{graphicx}
\usepackage{amsmath}
\usepackage{subfigure}
\usepackage{hyperref}
\usepackage{threeparttable, booktabs} 

\usepackage{hyperref}

\usepackage[accepted]{icml2018}

\icmltitlerunning{Cover Song Detection with Siamese Networks}

\begin{document}

\twocolumn[
\icmltitle{Towards Cover Song Detection with Siamese Convolutional Neural Networks}



\icmlsetsymbol{equal}{}

\begin{icmlauthorlist}
\icmlauthor{Marko Stamenovic}{equal,to}
\end{icmlauthorlist}

\icmlaffiliation{to}{Bose Corporation, Boston, MA, USA}

\icmlcorrespondingauthor{Marko Stamenovic}{\href{mailto:marko_stamenovic@bose.com}{marko{\textunderscore}stamenovic@bose.com}}

\icmlkeywords{Machine Learning, ICML}

\vskip 0.3in
]



\printAffiliationsAndNotice{} 

\begin{abstract}
A cover song, by definition, is a new performance or recording of a previously recorded, commercially released song. It may be by the original artist themselves or a different artist altogether and can vary from the original in unpredictable ways including key, arrangement, instrumentation, timbre and more. In this work we propose a novel approach to learning audio representations for the task of cover song detection. We train a neural architecture on tens of thousands of cover-song audio clips and test it on a held out set. We obtain a mean precision@1 of 65\% over mini-batches, ten times better than random guessing. Our results indicate that Siamese network configurations show promise for approaching the cover song identification problem. 
\end{abstract}

\section{Introduction}
\label{Introduction}

The proliferation of cheap digital media creation tools and free web based publishing platforms has led to an ever-expanding universe of audio-visual content. Although much of this content is original in nature, a stunning amount is cover material. For example, a recent search of a popular video sharing site for the term “Beatles cover” turned up 3.97 million matches. Over their entire career, The Beatles released a total of 257 songs. Applications of cover song detection include copyright infringement detection and content-based music recommendation among others.

Classical approaches to cover-song detection use feature extraction combined with a distance metric. Chroma features, which globally bin spectral semitone energy, are widely used as as input feature representations. Ellis \cite{Ellis_2006} used beat-synchronous chroma features and cross-correlation to detect cover songs. Ellis and Cotton \cite{Ellis_Cotton_2007} built on this work by tuning the cross correlation results and the tempo estimation. Serra et al. \cite{serra_2008a} used dynamic time warping to align a chroma-based features before similarity evaluation. More recently, Silva et al. \cite{DBLP:conf/ismir/SilvaYBK16} used an approach based on subsequence comparisons, Seetharaman and Rafii \cite{prem_rafii} proposed a method based on a two-dimensional Fourier transform of a binarized constant-Q spectrogram, and there have been a wide variety of approaches based on feature and sensor network fusion \cite{fusion2, DBLP:journals/corr/Tralie17}.

There is less literature with respect to machine learning approaches. Chang et al.  \cite{coversong_convnet} use a chroma self-similarity matrix as input to convolutional neural network and frame the problem as a classification task. Humphrey et al. \cite{HumphreyEtAl_2013_DataDrivAndDisc} use sparse-coding with two-dimensional Fourier magnitude coefficients extracted from chroma features.

In this work we propose a Siamese  convolutional network (convnet) based approach to cover song detection. First, we train the network to predict whether a pair of songs is a cover pair or not. We then use the trained network as a feature extractor to generate feature vectors which we use to calculate cover-song similarity.

Our approach differs from previous approaches to cover-song detection in that, outside of our initial time-spectral representation, the bulk of our feature extraction is done using machine learning rather than hand-tuned features. 

\section{Related Work}
\label{related work}

Siamese networks have been used since the early 1990s for learning similarity representations. The networks are referred to as ``Siamese'' networks because they consist of two identical sub-networks used to extract task-salient feature representations from the two input items to be compared. Baldi and Chauvin \cite{1993_siamese} propose Siamese network for fingerprint recognition. Bromley et al. \cite{bromley_lecun_siamese} describe such a network for verifying written signatures on a pen-input tablet. 

There has been much work done in recent years using CNN's on spectro-temporal audio representations for classification tasks \cite{hershey_cnn, AudioSet, keunwoo_ismir_17}. Siamese networks have been applied to audio tasks as well, including by Zhang and Duan \cite{yichi} for search by vocal imitation.

\section{Overview}
\label{overview}

\subsection{Dataset}

Our raw data consists of 64 Kbps 22.5 kHz MP3’s from 7-Digital preview clips of songs in the Second-Hand-Songs Dataset (SHS) \cite{Bertin-Mahieux2011} train set, which is a subset of the Million Song Dataset (MSD) \cite{msd}. Preview clips range from 30 to 60 seconds of the song pulled from random parts of the the track. The SHS contains a training set of 12,960 unique song clips, divided into 4,128 cliques, or version-groups of the same original song. For example there may be 5 versions of the same song, which would make up one clique. Overall, the dataset contains 24,986 cover-song pairs. During training we stochastically generate an equal amount of non-cover-song pairs by sampling songs from SHS which do not appear in the same clique. This amounts to 49,972 total song-pairs evenly split between cover-song pairs and non-cover-song pairs. We hold out 1000 song pairs as a validation set.

\subsection{Input Representation}

We use a time-frequency log-spectral representation of the audio based on the Constant Q Transform (CQT) \cite{cqt} which we extract using the Librosa python toolbox \cite{librosa}. The CQT is a transform with a logarithmic frequency resolution, mirroring the Western music scale and the human auditory perception of music. We use a time resolution corresponding to approximately 0.23 seconds per time frame and a frequency resolution of one semi-tone per frequency bin spanning 7 octaves from C1 (32Hz) to B7 (3951 Hz).

\subsection{Network Architecture and Hyperparameters}

Our network consists of a Siamese convolutional architecture shown in Figure 1. We train the model with a NVidia Titan X GPU for 5 epochs and a mini-batch size of 16 using the ADAM optimizer to update weights. To prevent overfitting, a dropout factor of 0.5 is used on the fully connected layers in addition to an L2 regularization lambda factor of 0.005.

\begin{figure}
\vskip 0.01 in
\begin{center}
\centerline{\includegraphics[width=0.5\columnwidth]{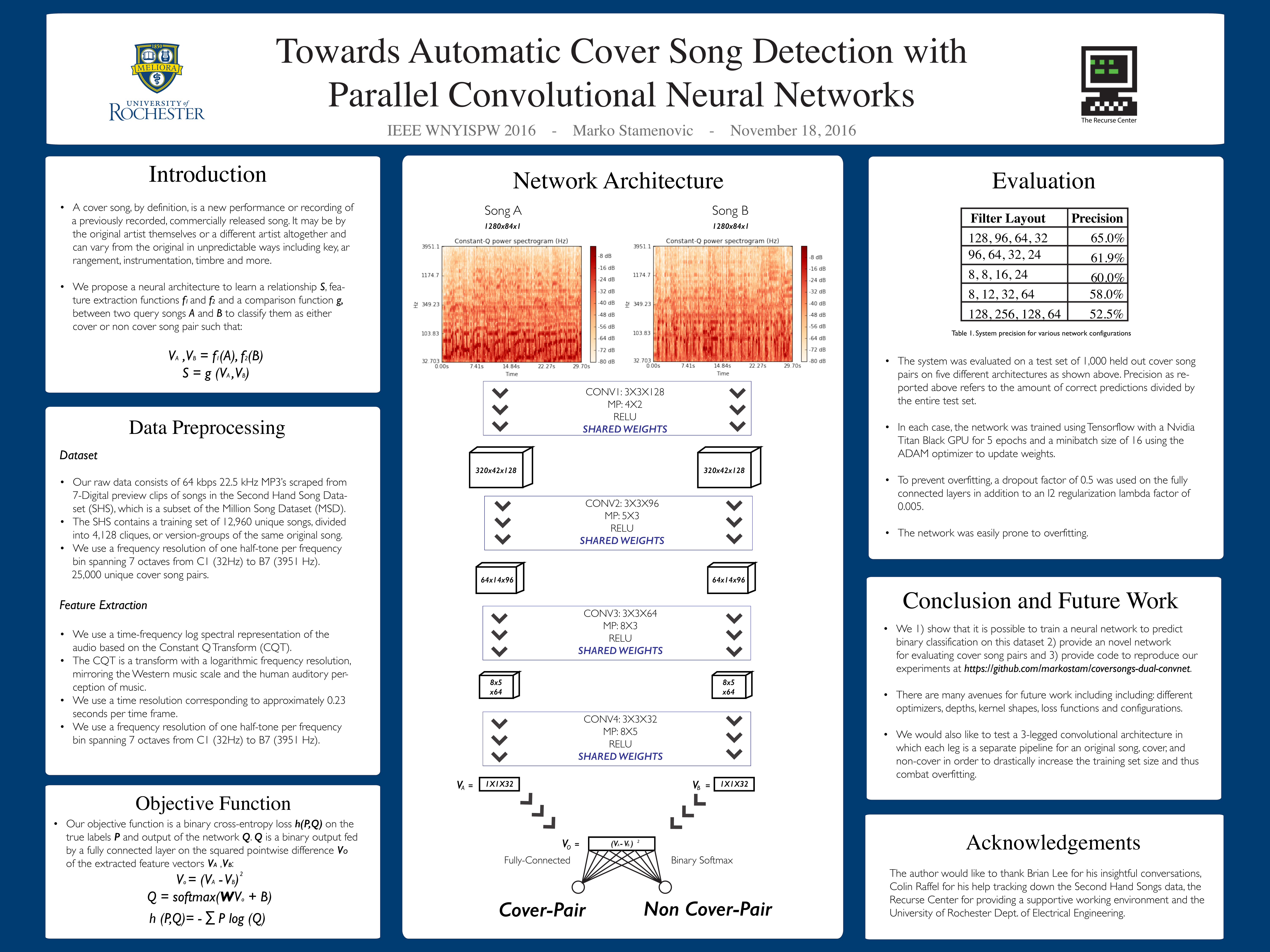}}
\caption{Network architecture and hyperparameters of the proposed algorithm.}
\label{icml-historical}
\end{center}
\vskip -0.01in
\end{figure}

\subsection{Objective Function}

To formalize the problem, our network is trying to learn a feature extraction function ${f}$ between two query songs $x_{a}$ and $x_{b}$ to classify them as either a cover or non cover-song pair. Let $v_{a}, v_{b} = {f}(x_{a}, x_{b})$, where $v_{a}, v_{b}$ are the output representations of $x_{a}$ and $x_{b}$, respectively, and ${f}$ is the convnet feature extractor. We then pass the output representations to a  comparison function ${p}$, shown below, where $\sigma$ is the sigmoidal activation function and $\alpha$ is an additional learned parameter weighting the importance of each index of ${p}$. This  comparison function computes the final similarity score between the output representations and fuses the Siamese network.

\begin{align*}
p(x_{a}, x_{b}) = \sigma\sum_{j}\alpha_{j}( v_{a}^{(j)} - v_{b}^{(j)})^2
\end{align*}

Our objective function is a binary cross-entropy loss $\mathcal{L} (x_{a}, x_{b})$ on the true labels ${y}(x_{a}, x_{b})$ and output of the network ${p}$, where we assume ${y}(x_{a}, x_{b}) = 1$ for each cover-song pair and $\mathbf{y}(x_{a}, x_{b}) = 0$ for each non cover-song pair.
\begin{align*}
\mathcal{L} (x_{a}, x_{b}) = {y}(x_{a}, x_{b})  \log {p}(x_{a}, x_{b}) + \\
(1 - {y}(x_{a}, x_{b}))  \log  (1-{p}(x_{a}, x_{b}))
\end{align*}

\subsection{Evaluation}

Performance of the network is evaluated by extracting features from the penultimate layer of the network and measuring cosine distance. We use a metric of mean precision at one (Prec@1) over a mini-batch size of 16. If a song is nearer to its cover pair than any other song in the mini-batch, we score it as correct and vice-versa. 

\section{Results}
Our best configuration yields a a Prec@1 score of 65.0\% over a mini-batch of 16. Since we evaluate over a mini-batch size of 16, random guessing would yield a Prec@1 of 0.0625\%. 

We compare our results on the SHS subset to various published results on the US Pop Music Collection Cover Song dataset (Mixed Collection) from MIREX \cite{Downie2010} in Table 1. This dataset consists of 30 cliques, each with 11 different versions, for a total of 330 songs. 

We evaluate a variety of convolutional network structures inspired by \cite{choi16} by varying the amount of filters per layer and whether the amount of filters per layer is increasing or decreasing proportional to network depth. We find that in general for this task starting with a large amount of filters at the input layer and gradually decreasing yields better performance. We also find that the network tends to overfit quickly, which is why we employ a relatively high rate of dropout and L2 regularization.

\section{Conclusion and Future Work}

In this work we show a novel machine-learning based approach to the cover song detection task, evaluate different network configurations to find a best suited architecture and provide code to reproduce our experiments \footnote{\url{https://github.com/markostam/coversongs-dual-convnet}}. In future work we would like to like to better evaluate the network to more closely match work done in the cover song detection space. We are also interested in exploring network structures such as attention that can map similarities between different temporal pieces of audio, triplet networks, and data augmentation. 
\begin{table}
\caption{Mean precision at one; our system compared to other published results. Note that ours is evaluated and averaged over mini-batches of 16 on the SHS while the other results use the MIREX Mixed Collection for evaluation.}
\label{results-table}
\vskip 0.15in
\begin{center}
\begin{small}
\begin{sc}
\begin{tabular}{lcccr}
\toprule
System & Prec@1 \\
\midrule
\cite{Ellis_Cotton_2007}    & 49.7 \\
\cite{Bello07audio-basedcover} & 45.8 \\
\cite{serra_2008a}  & 73.3 \\
\cite{fusion2}       & \textbf{76.6} \\
Ours\tnote{*}  & 65.0 \\
\bottomrule
\end{tabular}
\end{sc}
\end{small}
\end{center}
\vskip -0.1in
\end{table}
\label{results}


\bibliography{example_paper}

\begin{thebibliography}{22}
\providecommand{\natexlab}[1]{#1}
\providecommand{\url}[1]{\texttt{#1}}
\expandafter\ifx\csname urlstyle\endcsname\relax
  \providecommand{\doi}[1]{doi: #1}\else
  \providecommand{\doi}{doi: \begingroup \urlstyle{rm}\Url}\fi

\bibitem[Baldi \& Chauvin(1993)Baldi and Chauvin]{1993_siamese}
Baldi, Pierre and Chauvin, Yves.
\newblock Neural networks for fingerprint recognition.
\newblock \emph{Neural Comput.}, 5\penalty0 (3):\penalty0 402--418, May 1993.
\newblock ISSN 0899-7667.
\newblock \doi{10.1162/neco.1993.5.3.402}.
\newblock URL \url{http://dx.doi.org/10.1162/neco.1993.5.3.402}.

\bibitem[Bello(2007)]{Bello07audio-basedcover}
Bello, Juan~Pablo.
\newblock Audio-based cover song retrieval using approximate chord sequences:
  testing shifts, gaps, swaps and beats.
\newblock In \emph{Int. Symp. on Music Information Retrieval (ISMIR}, pp.\
  239--244, 2007.

\bibitem[Bertin-Mahieux et~al.(2011{\natexlab{a}})Bertin-Mahieux, Ellis,
  Whitman, and Lamere]{Bertin-Mahieux2011}
Bertin-Mahieux, Thierry, Ellis, Daniel~P.W., Whitman, Brian, and Lamere, Paul.
\newblock The million song dataset.
\newblock In \emph{{Proceedings of the 12th International Conference on Music
  Information Retrieval ({ISMIR} 2011)}}, 2011{\natexlab{a}}.

\bibitem[Bertin-Mahieux et~al.(2011{\natexlab{b}})Bertin-Mahieux, Ellis,
  Whitman, and Lamere]{msd}
Bertin-Mahieux, Thierry, Ellis, Daniel~P.W., Whitman, Brian, and Lamere, Paul.
\newblock The million song dataset.
\newblock In \emph{{Proceedings of the 12th International Conference on Music
  Information Retrieval ({ISMIR} 2011)}}, 2011{\natexlab{b}}.

\bibitem[Bromley et~al.(1994)Bromley, Guyon, LeCun, S\"{a}ckinger, and
  Shah]{bromley_lecun_siamese}
Bromley, Jane, Guyon, Isabelle, LeCun, Yann, S\"{a}ckinger, Eduard, and Shah,
  Roopak.
\newblock Signature verification using a "siamese" time delay neural network.
\newblock In Cowan, J.~D., Tesauro, G., and Alspector, J. (eds.),
  \emph{Advances in Neural Information Processing Systems 6}, pp.\  737--744.
  Morgan-Kaufmann, 1994.
\newblock URL
  \url{http://papers.nips.cc/paper/769-signature-verification-using-a-siamese-time-delay-neural-network.pdf}.

\bibitem[Brown et~al.(1992)Brown, Puckette, of~Technology. Media
  Laboratory.~Vision, and Group]{cqt}
Brown, J.C., Puckette, M., of~Technology. Media Laboratory.~Vision,
  Massachusetts~Institute, and Group, Modeling.
\newblock \emph{An Efficient Algorithm for the Calculation of a Constant Q
  Transform}.
\newblock M.I.T. Media Lab Vision and Modeling Group technical report. Vision
  and Modeling Group, Media Laboratory, Massachusetts Institute of Technology,
  1992.
\newblock URL \url{https://books.google.com/books?id=rao7HQAACAAJ}.

\bibitem[Chang et~al.(2017)Chang, Lee, Choe, and Lee]{coversong_convnet}
Chang, Sungkyun, Lee, Juheon, Choe, Sang~Keun, and Lee, Kyogu.
\newblock Audio cover song identification using convolutional neural network.
\newblock \emph{CoRR}, abs/1712.00166, 2017.
\newblock URL \url{http://arxiv.org/abs/1712.00166}.

\bibitem[Chen et~al.(2017)Chen, Li, and Xiao]{fusion2}
Chen, Ning, Li, Wei, and Xiao, Haidong.
\newblock Fusing similarity functions for cover song identification.
\newblock 77, 02 2017.

\bibitem[Choi et~al.(2016)Choi, Fazekas, and Sandler]{choi16}
Choi, Keunwoo, Fazekas, George, and Sandler, Mark~B.
\newblock Automatic tagging using deep convolutional neural networks.
\newblock \emph{CoRR}, abs/1606.00298, 2016.
\newblock URL \url{http://arxiv.org/abs/1606.00298}.

\bibitem[Choi et~al.(2017)Choi, Fazekas, Sandler, and Cho]{keunwoo_ismir_17}
Choi, Keunwoo, Fazekas, Gy{\"{o}}rgy, Sandler, Mark~B., and Cho, Kyunghyun.
\newblock Transfer learning for music classification and regression tasks.
\newblock \emph{CoRR}, abs/1703.09179, 2017.
\newblock URL \url{http://arxiv.org/abs/1703.09179}.

\bibitem[Downie et~al.(2010)Downie, Ehmann, Bay, and Jones]{Downie2010}
Downie, J.~Stephen, Ehmann, Andreas~F., Bay, Mert, and Jones, M.~Cameron.
\newblock \emph{The Music Information Retrieval Evaluation eXchange: Some
  Observations and Insights}, pp.\  93--115.
\newblock Springer Berlin Heidelberg, Berlin, Heidelberg, 2010.
\newblock ISBN 978-3-642-11674-2.
\newblock \doi{10.1007/978-3-642-11674-2_5}.
\newblock URL \url{https://doi.org/10.1007/978-3-642-11674-2_5}.

\bibitem[Ellis(2006)]{Ellis_2006}
Ellis, Daniel P.~W.
\newblock Identifying 'cover songs' with beat-synchronous chroma features.
\newblock In \emph{Music Information Retrieval Evaluation eXchange 2006}, 2006.

\bibitem[Ellis \& Cotton(2007)Ellis and Cotton]{Ellis_Cotton_2007}
Ellis, Daniel P.~W. and Cotton, Courtenay~V.
\newblock The 2007 labrosa cover song detection system.
\newblock In \emph{Music Information Retrieval Evaluation eXchange 2007}, 2007.

\bibitem[Gemmeke et~al.(2017)Gemmeke, Ellis, Freedman, Jansen, Lawrence, Moore,
  Plakal, and Ritter]{AudioSet}
Gemmeke, Jort~F., Ellis, Daniel P.~W., Freedman, Dylan, Jansen, Aren, Lawrence,
  Wade, Moore, R.~Channing, Plakal, Manoj, and Ritter, Marvin.
\newblock Audio set: An ontology and human-labeled dataset for audio events.
\newblock In \emph{Proc. IEEE ICASSP 2017}, New Orleans, LA, 2017.

\bibitem[Hershey et~al.(2017)Hershey, Chaudhuri, Ellis, Gemmeke, Jansen, Moore,
  Plakal, Platt, Saurous, Seybold, Slaney, Weiss, and Wilson]{hershey_cnn}
Hershey, Shawn, Chaudhuri, Sourish, Ellis, Daniel P.~W., Gemmeke, Jort~F.,
  Jansen, Aren, Moore, Channing, Plakal, Manoj, Platt, Devin, Saurous, Rif~A.,
  Seybold, Bryan, Slaney, Malcolm, Weiss, Ron, and Wilson, Kevin.
\newblock Cnn architectures for large-scale audio classification.
\newblock In \emph{International Conference on Acoustics, Speech and Signal
  Processing (ICASSP)}. 2017.
\newblock URL \url{https://arxiv.org/abs/1609.09430}.

\bibitem[Humphrey et~al.(2013)Humphrey, Nieto, and
  Bello]{HumphreyEtAl_2013_DataDrivAndDisc}
Humphrey, Eric~J., Nieto, Oriol, and Bello, Juan~P.
\newblock Data driven and discriminative projections for large-scale cover song
  identification.
\newblock In \emph{Proceedings of the 14th International Society for Music
  Information Retrieval Conference}, November 4-8 2013.
\newblock
  \url{http://www.ppgia.pucpr.br/ismir2013/wp-content/uploads/2013/09/246_Paper.pdf}.

\bibitem[McFee et~al.(2017)McFee, McVicar, Nieto, Balke, Thome, Liang,
  Battenberg, Moore, Bittner, Yamamoto, Ellis, Stoter, Repetto, Waloschek,
  Carr, Kranzler, Choi, Viktorin, Santos, Holovaty, Pimenta, and Lee]{librosa}
McFee, Brian, McVicar, Matt, Nieto, Oriol, Balke, Stefan, Thome, Carl, Liang,
  Dawen, Battenberg, Eric, Moore, Josh, Bittner, Rachel, Yamamoto, Ryuichi,
  Ellis, Dan, Stoter, Fabian-Robert, Repetto, Douglas, Waloschek, Simon, Carr,
  CJ, Kranzler, Seth, Choi, Keunwoo, Viktorin, Petr, Santos, Joao~Felipe,
  Holovaty, Adrian, Pimenta, Waldir, and Lee, Hojin.
\newblock librosa 0.5.0, February 2017.
\newblock URL \url{https://doi.org/10.5281/zenodo.293021}.

\bibitem[Seetharaman \& Rafii(2017)Seetharaman and Rafii]{prem_rafii}
Seetharaman, P. and Rafii, Z.
\newblock Cover song identification with 2d fourier transform sequences.
\newblock In \emph{2017 IEEE International Conference on Acoustics, Speech and
  Signal Processing (ICASSP)}, pp.\  616--620, March 2017.
\newblock \doi{10.1109/ICASSP.2017.7952229}.

\bibitem[Serr{\`a} et~al.(2008)Serr{\`a}, G{\'o}mez, Herrera, and
  Serra]{serra_2008a}
Serr{\`a}, Joan, G{\'o}mez, Emilia, Herrera, Perfecto, and Serra, Xavier.
\newblock Chroma binary similarity and local alignment applied to cover song
  identification.
\newblock \emph{IEEE Transactions on Audio, Speech and Language Processing},
  16:\penalty0 1138--1151, 08/2008 2008.
\newblock ISSN 1558-7916.
\newblock \doi{10.1109/TASL.2008.924595}.
\newblock URL \url{files/publications/jserraTSALP08.pdf}.

\bibitem[Silva et~al.(2016)Silva, Yeh, Batista, and
  Keogh]{DBLP:conf/ismir/SilvaYBK16}
Silva, Diego~Furtado, Yeh, Chin{-}Chia~Michael, Batista, Gustavo E. A. P.~A.,
  and Keogh, Eamonn~J.
\newblock Simple: Assessing music similarity using subsequences joins.
\newblock In \emph{Proceedings of the 17th International Society for Music
  Information Retrieval Conference, {ISMIR} 2016, New York City, United States,
  August 7-11, 2016}, pp.\  23--29, 2016.
\newblock URL
  \url{https://wp.nyu.edu/ismir2016/wp-content/uploads/sites/2294/2016/07/099_Paper.pdf}.

\bibitem[Tralie(2017)]{DBLP:journals/corr/Tralie17}
Tralie, Christopher~J.
\newblock Early {MFCC} and {HPCP} fusion for robust cover song identification.
\newblock \emph{CoRR}, abs/1707.04680, 2017.
\newblock URL \url{http://arxiv.org/abs/1707.04680}.

\bibitem[Zhang \& Duan(2017)Zhang and Duan]{yichi}
Zhang, Y. and Duan, Z.
\newblock Iminet: Convolutional semi-siamese networks for sound search by vocal
  imitation.
\newblock In \emph{2017 IEEE Workshop on Applications of Signal Processing to
  Audio and Acoustics (WASPAA)}, pp.\  304--308, Oct 2017.
\newblock \doi{10.1109/WASPAA.2017.8170044}.

\end{thebibliography}
\bibliographystyle{icml2018}

\end{document}